\documentclass[prl,twocolumn,superscriptaddress,showpacs]{revtex4}

\usepackage{amsmath,graphicx,latexsym,bm}
\usepackage{psfrag,color}

\begin{document}

\title{Surface control of flexoelectricity}

\author{Massimiliano Stengel}
\affiliation{ICREA - Instituci\'o Catalana de Recerca i Estudis Avan\c{c}ats, 08010 Barcelona, Spain}
\affiliation{Institut de Ci\`encia de Materials de Barcelona 
(ICMAB-CSIC), Campus UAB, 08193 Bellaterra, Spain}

\date{\today}

\begin{abstract}
The polarization response of a material to a strain gradient, known as flexoelectricity,
holds great promise for novel electromechanical applications. Despite considerable
recent progress, however, the effect remains poorly understood. 
From both the fundamental and practical viewpoints, it is of crucial 
importance to know whether the coupling coefficients are primarily governed by 
the properties of the bulk material or by the details of the sample surface.
Here we provide, by means of first-principles calculations, 
quantitative evidence supporting the latter scenario.
In particular, we demonstrate that a SrTiO$_3$ film can yield a positive
or negative flexoelectric voltage depending on its surface termination.
This result points to a full control of the flexoelectric effect via 
surface/interface engineering, opening exciting new avenues for device design.
\end{abstract}

\pacs{71.15.-m,77.65.-j}

\maketitle

Flexoelectricity is a universal property of all insulators, whereby a
macroscopic electrical polarization is generated in response to an inhomogeneous
mechanical strain.~\cite{Tagantsev}
The recent surge of interest in this phenomenon~\cite{pavlo_review} has come with 
the realization that strain gradients can be huge at the nanoscale, and generate a 
large enough polarization to rival conventional piezoelectricity.~\cite{Cross} 
Long regarded as a drawback in the operation of thin-film devices
(e.g. ferroelectric memories~\cite{Gustau-05}, or light-emitting diodes in
foldable electronics~\cite{iled}), strain gradients are now being increasingly recognized
as a rich playground for exploring new, potentially useful, functionalities~\cite{Gustau1,Gustau2}.
Future progress towards practical applications crucially relies on identifying the 
microscopic mechanism that are most effective at delivering a large electrical response, and on 
harnessing them via specific materials-design rules.
Unfortunately, such a fundamental knowledge is currently very limited.

A central (yet vastly unexplored) question concerns the 
role played by the sample surfaces.
Both symmetry arguments~\cite{Taga_finite} and quantum-mechanical theory~\cite{Hong-11,artgr} 
predict their contribution to be comparable to that of the bulk, regardless of sample size.
This fact, in principle, calls for a substantial revision of the currently established
device design strategies, where only bulk electromechanical properties (such as 
piezoelectricity~\cite{Martin} and/or electrostriction) are typically taken into 
account.~\cite{review_semi}
In practice, however, there are currently little or no indications
on the magnitude of the aforementioned surface effects, 
nor on their microscopic physical nature, so the necessity for their explicit
inclusion in the models remains open to debate.
It appears unlikely that such indications will emerge from the experiments alone, 
at least in the near future: as the bulk and surface contributions scale identically
as a function of sample size (unlike other interface-related phenomena in oxides, e.g. 
the dielectric ``dead'' layer~\cite{nature_2006,nature_mat}),
they appear difficult or impossible to disentangle by purely electrical means.
In this context, theoretical modeling can be of great help.

First-principles electronic-structure methods are, in principle, ideally suited 
to shed some light on the above issues, thanks to their unbiased predictive power.
Indeed, building on the work of Resta~\cite{Resta-10}, Hong and Vanderbilt~\cite{Hong-11,Hong-13}
have recently devised a promising route to addressing the flexoelectric problem
at a fundamental quantum-mechanical level, and applied it to calculating the response 
properties of a number of bulk materials in the framework of density-functional theory (DFT).
However, even at the bulk level, a complete determination of the full flexoelectric tensor 
has not been achieved yet, as the electronic contribution to the 
transversal components is still missing~\cite{Hong-13}. Furthermore, the impact of surface effects was not
considered in refs.~\cite{Hong-11,Hong-13}, nor in any other first-principles study~\cite{mara/sha,Hong-10} 
reported to date.
Very recent advances promoted by the author~\cite{artlin,artgr} have now opened the
way to filling both gaps, by combining density-functional perturbation theory (DFPT) 
--the linear-response version of DFT-- with a covariant formulation of electrostatics
in the coordinate system of the deformed body.
Here we use such a methodology to study the flexoelectric response of 
SrTiO$_3$, arguably the most important material for 
applications, and the best known experimentally~\cite{Pavlo}.  
Our results, in addition to providing a complete physical picture of the effect, 
demonstrate that the surface indeed matters: by modifying the atomically thin 
termination layer, one can tune --and even reverse-- the 
voltage response of a macroscopically thick film.

\begin{figure}
\begin{flushright}
\includegraphics[width=3.3in,clip]{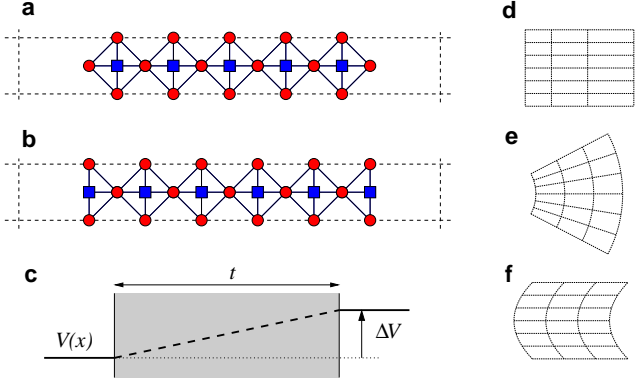} \\
\end{flushright}
\caption{ \label{fig1} Schematic illustration of the computational setup. Panels (a,b) show the 
supercell models of the SrO- (a) and TiO$_2$-terminated (b) SrTiO$_3$ slabs (Ti and O atoms are 
represented as squares and circles; Sr atoms are not shown). Panel (c) shows the induced 
open-circuit voltage, $\Delta V$, induced by a longitudinal ($\varepsilon_{xx,x}$, d), transversal 
($\varepsilon_{yy,x}$, e) or shear ($\varepsilon_{xy,y}$, f) strain-gradient deformation.}
\end{figure}

The flexoelectric performance of an insulating material 
can be conveniently quantified as the open-circuit voltage, $\Delta V$, that is linearly
induced by a strain-gradient deformation [see Fig.~\ref{fig1}(c)] in the limit of a large 
film thickness, $t$,
\begin{equation}
\varphi_{x \lambda, \beta \gamma} = \lim_{t \rightarrow \infty} \frac{1}{t} 
\frac{\partial \Delta V}{\partial \varepsilon_{\beta \gamma, \lambda}}.
\label{varphi1}
\end{equation}
Here $\varepsilon_{\beta \gamma, \lambda} = \partial \varepsilon_{\beta \gamma} / \partial r_\lambda$ 
is the gradient of the symmetric strain tensor ($\varepsilon_{\beta \gamma}$) along the Cartesian direction $r_\lambda$,
and $x$ indicates the direction normal to the surface;
the relevant components of $\varepsilon_{\beta \gamma, \lambda}$ in the context of this work are 
illustrated in Fig.~\ref{fig1}(d-f).
Remarkably, although $\varphi_{x \lambda, \beta \gamma}$ (usually referred to as ``flexocoupling'' coefficient)
is a macroscopic response property of the system, it is known~\cite{artgr} to contain \emph{both} bulk- and
surface-specific contributions,
\begin{equation}
\varphi_{x \lambda, \beta \gamma} = \varphi_{x \lambda, \beta \gamma}^{\rm bulk} +
\varphi_{x \lambda, \beta \gamma}^{\rm surf}.
\label{varphi2}
\end{equation}
The former term is given by 
$\varphi_{x \lambda, \beta \gamma}^{\rm bulk}=\mu_{x \lambda, \beta \gamma}^{\rm bulk} / (\epsilon_0 \epsilon_{xx})$,
where $\bm{\mu}^{\rm bulk}$ 
is the bulk flexoelectric tensor, $\bm{\epsilon}$ is the macroscopic
dielectric tensor and $\epsilon_0$ is the vacuum permittivity.
The latter term, $\varphi_{x \lambda, \beta \gamma}^{\rm surf}$, originates from 
surface piezoelectric effects,~\cite{Taga_finite,artgr} which are present in
any material regardless of crystal symmetry. 

While techniques for calculating both quantities in a first-principles context
have recently been proposed,~\cite{artlin,artgr,Hong-13} the determination of 
the bulk flexoelectric tensor remains challenging, particularly concerning 
the purely electronic contributions.
In fact, directly calculating the bulk polarization response to a strain 
gradient would require access to the microscopic current density~\cite{artlin,Hong-13} 
induced by a deformation, whose code implementation is not available yet.
To overcome this methodological obstacle (and solve for the missing transversal 
components of $\varphi^{\rm bulk}$) we shall compute, rather than the
polarization response of the \emph{bulk}, the internal electric field response 
of a \emph{slab}. 
The advantage of the latter approach is that it can be carried out with the
sole knowledge of the first-order charge density.~\cite{artlin}
%
%
For the sake of computational convenience, we shall initially focus on ``frozen-ion'' 
(in the sense specified in ref.~\cite{artgr}) deformation of ``truncated-bulk'' 
slabs, i.e., with the unperturbed atoms placed at their ideal lattice sites.
The impact of full ionic relaxation, which is essential for a quantitative analysis
of the flexoelectric effect, is uncomplicated to calculate once the electronic
contributions to the bulk tensor are known, and will be dealt with in a later part 
of this work.
%

We consider the supercell models illustrated in Fig.~\ref{fig1}(a-b) (i.e. 
periodically repeated sequences of symmetrically terminated SrTiO$_3$ slabs and vacuum layers), 
and proceed as follows.
First, we calculate how the microscopic charge density of the supercell, 
$\rho({\bf r})$, responds to a selected set of 
long-wavelength acoustic phonons, by using DFPT as implemented in the ABINIT~\cite{abinit,gonze}
package.
Next, we perform a Taylor expansion (in the wavevector ${\bf q}$) of such density 
response functions, along the lines described in refs.~\cite{artlin,artgr}. 
This analysis readily yields the response to a macroscopic strain gradient 
in the curvilinear coordinate system of the deformed crystal lattice;~\cite{artgr}
in particular, one has
\begin{eqnarray}
\frac{\partial \rho({\bf r})}{\partial \varepsilon_{\beta \gamma,\lambda}} &=& 
r_\lambda \rho^{\rm U}_{\beta \gamma}({\bf r}) +
\rho^{\rm G}_{\lambda, \beta \gamma}({\bf r}), 
\label{fhat}
\end{eqnarray}
where 
$\rho^{\rm U,G}$ are cell-periodic functions 
[$\rho^{\rm U}=\partial \rho({\bf r}) / \partial \varepsilon_{\beta \gamma}$ 
describes the response to a uniform strain, while $\rho^{\rm G}$ is
the additional contribution that is due to the gradient].
Finally, we use $\rho^{\rm U}$ and $\rho^{\rm G}$ to calculate the induced electric field.
In the curvilinear frame, the first-order ${\bf E}$ is related to the first-order $\rho$ 
via the modified~\cite{artgr} Gauss's law,
\begin{equation}
\nabla \cdot \left( \frac{\partial {\bf E}({\bf r})}{\partial \varepsilon_{\beta \gamma,\lambda}} + 
r_\lambda {\bf E}^{\rm met}_{\beta \gamma}({\bf r}) \right) = 
\frac{1}{\epsilon_0} \frac{\partial \rho({\bf r})}{\partial \varepsilon_{\beta \gamma,\lambda}}
\label{metric}
\end{equation}
where the metric contribution ${\bf E}^{\rm met}$ depends on the unperturbed electric field ${\bf E}$ as
\begin{equation}
E^{\rm met}_{\alpha,\beta \gamma} = \delta_{\beta \gamma} E_\alpha - \delta_{\gamma \alpha} E_\beta - \delta_{\alpha \beta} E_\gamma.
\end{equation}
As the electric field is related to the potential by 
${\bf E}({\bf r}) = - \nabla V({\bf r})$, 
knowledge of the former (and of its first-order variation, 
$\partial {\bf E}({\bf r}) / \partial \varepsilon_{\beta \gamma,\lambda}$) then yields 
the linear variation of the latter, 
$\partial V({\bf r}) / \partial \varepsilon_{\beta \gamma,\lambda}$,
and ultimately the sought-after values of the flexocoupling coefficients, $\varphi$.

In practice, to solve Eq.~(\ref{metric}), it is convenient to work with the
``macroscopic averages''~\cite{Baldereschi-88,Junquera-07} of the ${\bf E}$ and $\rho$
response functions, where the oscillations that occur on the scale of the interatomic 
spacings have been appropriately filtered out.
This procedure has two advantages: first, it allows one to identify the 
relevant electrical properties of the system in a macroscopic context 
(e.g. internal fields, surface potential offsets, etc.); second, 
it facilitates the implementation of the Poisson solver
by making the problem one-dimensional. 
In particular, in close analogy to Eq.~(\ref{fhat}), one can write the 
normal ($x$) component of the macroscopically averaged ${\bf E}$-field response as
\begin{eqnarray}
\frac{\partial E_x(x)}{\partial \varepsilon_{\beta \beta,x}} &=& 
x E^{\rm U}_{x,\beta \beta}(x) +
E^{\rm G}_{xx,\beta \beta}(x), \label{ex1} \\ 
\frac{\partial E_x(x)}{\partial \varepsilon_{xy,y}} &=& 
E^{\rm G}_{xy,xy}(x), \label{ex2} 
\end{eqnarray}
where Eq.~(\ref{ex1}) refers to either the longitudinal or 
transversal case ($\beta=x,y$), and Eq.~(\ref{ex2}) concerns a shear deformation.
Note that the parallel components of the induced ${\bf E}$-field vanish, 
hence the exclusive focus on $E_x$.
Note also the absence of the uniform-strain contribution in Eq.~(\ref{ex2}):
$E^{\rm U}_{x,x y}(x)$ vanishes identically in a centrosymmetric slab.

\begin{figure}
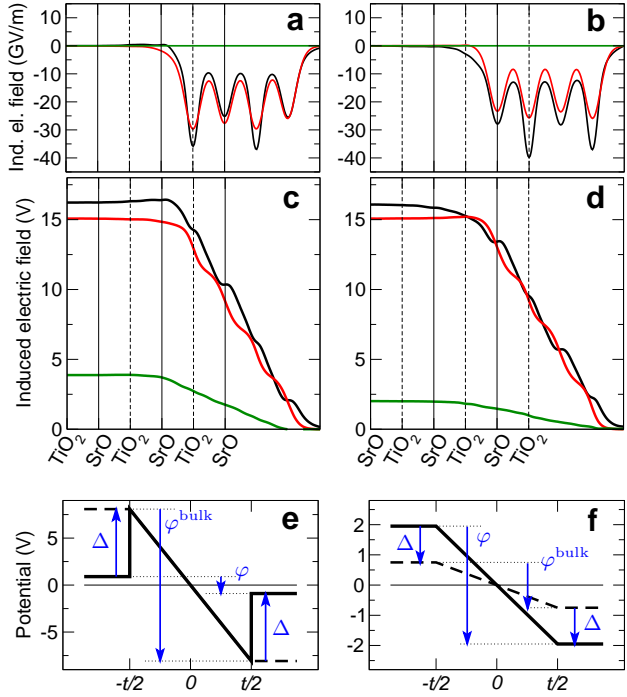

\psfrag{FSU}[][][1]{ \textcolor{blue}{ $\frac{\varphi^{\rm surf}}{2}$ } }
\psfrag{FSL}[b][b][0.9]{\textcolor{blue}{$\varphi$}}
\psfrag{FBU}[b][b][0.9]{\textcolor{blue}{$\varphi^{\rm bulk}$}}
\psfrag{DE}[b][b][0.9]{\textcolor{blue}{$\Delta$}}
\begin{flushright}
\includegraphics[width=3.24in,clip]{figure_ef2_SI.eps} \\
\includegraphics[width=3.2in,clip]{figure_efield.eps} \\
\vspace{10pt}
\includegraphics[width=3.2in,clip]{potentials2.eps}
\end{flushright}
\caption{ \label{efields} Electric field and potential response to mechanical deformations. 
The $E^{\rm U}_x$ (a-b) and $E^{\rm G}_x$ (c-d) response functions are shown for a SrO- (a,c)
and TiO$_2$-terminated (b,d) slab. Black, red and green curves refer to longitudinal, transversal
and shear deformations, respectively. The location of the SrO (dashed) and TiO$_2$ (solid) atomic 
layers is indicated by vertical lines (only half of the symmetric slab is shown). 
Panels (e-f) show the electrostatic potential that would be 
induced in a macroscopic SrO-terminated slab of thickness $t$ when subjected to a strain gradient 
of hypotetical magnitude $1/t$. The longitudinal (e) and shear (f) cases are shown, illustrating
the qualitative difference in the response. Dashed and solid lines refer to the bulk and total 
contribution, respectively; $\Delta= \varphi^{\rm surf} / 2$.}
\end{figure}

In Fig.~\ref{efields}(a-d) we plot the calculated $E_x^{\rm U,G}(x)$, 
corresponding to either a SrO- or a TiO$_2$-terminated slab and
to each of the three types of strain gradients shown in Fig.~\ref{fig1}(d-f). 
(We refer the reader to Supplementary Note 1 for the details of the 
computational methodology and parameters.)
As anticipated in Eqs.~(\ref{ex1}) and~(\ref{ex2}), there is an important qualitative
difference between the longitudinal or transversal response, where the
strain gradient is oriented along the surface normal, and the shear response, 
where it is directed in plane. 

In the former two cases, $E_{x,\beta \beta}^{\rm U}(x)$ is 
roughly uniform and negative (the oscillations are irrelevant on a 
macroscopic scale) in a thin region surrounding the surface layer.
This is consistent with the expected behavior of the electrostatic 
potential upon uniform deformation of the slab: the unperturbed $V(x)$ is a 
symmetric potential well, whose depth, $\phi$, is modified by a diagonal strain
component, $\varepsilon_{\beta \beta}$.
Such a dependence of $\phi$ on the strain corresponds precisely to 
the surface contribution to the flexocoupling coefficient,
\begin{equation}
\varphi^{\rm surf}_{xx,\beta \beta} = 
\frac{d \phi}{d \epsilon_{\beta \beta}} = -\int_{0}^{+\infty} dx \, E_{x,\beta \beta}^{\rm U}(x).
\label{dphideps}
\end{equation}
The functions $E_{x x,\beta \beta}^{\rm G}(x)$, describing the genuine
strain-gradient effects, display a capacitor-like behavior: the field
is uniform inside the film and zero outside, consistent with the open-circuit 
electrical boundary conditions (EBC) that were enforced.
Interestingly, the internal field, $E^{\rm slab}_{x x,\beta \beta}=E_{xx,\beta \beta}^{\rm G}(x=0)$, appears 
to be \emph{independent} of the surface termination.
This is not a coincidence: the open-circuit 
flexoelectric field in the longitudinal and transversal case is a bulk property 
of the material, and relates to the bulk flexocoupling coefficients~\cite{artlin,artgr} as 
\begin{equation}
\varphi^{\rm bulk}_{xx,\beta \beta} = -E^{\rm slab}_{xx,\beta \beta}. 
\label{eslab}
\end{equation}
Thus, for a strain gradient of the type $\varepsilon_{\beta \beta,x}$, the analysis
of the ${\bf E}$-response of the deformed slab yields complete information on both
surface and bulk contributions to the flexoelectric effect [their respective
impact on the electrostatic potential of a macroscopic film is illustrated in
Fig.~\ref{efields}(e)]. 
Most importantly, we have thereby gained access to the transversal component of
the bulk flexocoupling coefficient, $\varphi^{\rm bulk}_{xx,yy}$, whose 
computation has eluded earlier first-principles attempts.


\begin{table}
\begin{center}
\begin{ruledtabular}
\begin{tabular}{crrrrr}
&  \multicolumn{1}{c}{$\varphi^{\rm bulk}$} & \multicolumn{2}{c}{$\varphi^{\rm surf}$} &  \multicolumn{2}{c}{$\varphi$ (total)} \\    
      &      & \multicolumn{1}{c}{SrO} &  \multicolumn{1}{c}{TiO$_2$} 
                                  & \multicolumn{1}{c}{SrO} &  \multicolumn{1}{c}{TiO$_2$} \\
\hline                                  
$xx,xx$ (L)  & $-$16.153  &    14.356    &     16.948  &   $-1$.797   &     0.795 \\
$xx,yy$ (T)  & $-$15.075  &    15.683    &     12.447  &    0.608   &     $-2$.628 \\
$xy,xy$ (S)  &  $-$1.495  &  $-2$.382    &   $-$0.515  &   $-3$.877   &    $-2$.010
\end{tabular}
\end{ruledtabular}
\end{center}
\caption{Frozen-ion flexocoupling coefficients of a truncated-bulk SrTiO$_3$ slab. 
To compute $\varphi^{\rm bulk}$ we combined the two quantities that we extracted from the
bulk calculations, $\varphi^{\rm bulk}_{\rm L1}=-16.15$ V and $\varphi^{\rm bulk}_{\rm L2}=-18.07$ V
(these are in excellent agreement with the values reported by Hong and Vanderbilt~\cite{Hong-13}: 
$\varphi^{\rm bulk}_{\rm L1}=-16.25$ V and $\varphi^{\rm bulk}_{\rm L2}=-18.17$ V), 
with the value of $E^{\rm slab}_{xx,yy} = 15.08$ V, which we extracted from Fig.~\ref{efields}(c-d). 
The other values have been obtained as explained in the text. (L), (T) and (S) stands for longitudinal,
transversal and shear, respectively.
Volt units are used throughout. \label{tab1}}
\end{table}

In the shear case, the flexoelectric field depends on both 
bulk and surface-specific properties,~\cite{artgr} and is therefore
termination-dependent [see Fig.~\ref{efields}(f)]; from the electric field response 
functions of Fig.~\ref{efields}(a-d) we can thus only extract the \emph{total} 
flexocoupling coefficient of the slab, $\varphi_{x y,xy} = -E^{\rm slab}_{xy,xy}$.
To separate $\varphi_{x y,xy}$ into bulk and surface terms it suffices, however, 
to complement the above data with a calculation of bulk SrTiO$_3$. (Details are
reported in Supplementary Note 2.)
The latter, in particular, yields two additional response quantities,~\cite{Hong-13}
\begin{eqnarray}
\varphi^{\rm bulk}_{\rm L1} &=& \varphi^{\rm bulk}_{xx,xx}, \label{phil1} \\
\varphi^{\rm bulk}_{\rm L2} &=& \varphi^{\rm bulk}_{xx,yy} + 2 \varphi^{\rm bulk}_{xy,xy}.
\label{phil2}
\end{eqnarray}
Eq.~(\ref{phil1}) constitutes a useful consistency check of the methodology,
as $\varphi^{\rm bulk}_{\rm L1}$ is redundant with the already calculated 
value of $\varphi^{\rm bulk}_{xx,xx}$. 
Eq.~(\ref{phil2}), on the other hand, yields the sought-after value 
of $\varphi^{\rm bulk}_{xy,xy}$ since we already know $\varphi^{\rm bulk}_{xx,yy}$ 
from the slab calculations.
Finally, we use $\varphi_{xy,xy} = -E^{\rm slab}_{xy,xy}$ to infer 
$\varphi^{\rm surf}_{x y, x y}=\varphi_{xy,xy} - \varphi^{\rm bulk}_{xy,xy}$.

Our results for the bulk, surface, and total flexocoupling 
coefficients of the truncated-bulk, frozen-ion deformation of a SrTiO$_3$ 
slab are summarized in Table~\ref{tab1}.
At the bulk level, it is interesting to note the relatively small magnitude 
of the shear coefficients, $\varphi^{\rm bulk}_{xy,xy}$ and $\varphi^{\rm surf}_{xy,xy}$,
compared to both the longitudinal and the transversal ones. 
Meanwhile, in the latter two cases there is a large cancellation 
between bulk and surface terms;
as a result, the values of the total flexocoupling coefficients, $\varphi$, 
are all comparable in magnitude.
This fact can be rationalized by observing that the linear
response to atomic displacements, in a ionic (or partially ionic) solid, is largely 
dominated by the rigid displacement of an approximately spherical charge
density distribution surrounding each atom.
The spherical contribution, which is typically large and negative,~\cite{Hong-11} 
shows up in  $\varphi^{\rm bulk}_{xx,\beta \beta}$, and with opposite sign
in $\varphi^{\rm surf}_{xx,\beta \beta}$; in the shear case neither the bulk
nor the surface term are affected. (See Supplementary Note 1 of Ref.~\cite{artgr} and 
Figure S2 therein.)
Remarkably, the resulting values of $\varphi$ depend 
strongly on the details of the surface, and in some cases even have
opposite signs in the SrO- and TiO$_2$-terminated slabs.
Such a conclusion, in fact, persists after we take into account the 
full relaxation of the atomic structure; we shall prove this point in 
the following paragraphs.


%





\begin{table}
\begin{center}
\begin{ruledtabular}
\begin{tabular}{crrrrr}
&  \multicolumn{1}{c}{$\varphi^{\rm bulk}$} & \multicolumn{2}{c}{$\varphi^{\rm surf}$} &  \multicolumn{2}{c}{$\varphi$ (total)} \\    
&      & \multicolumn{1}{c}{SrO} &  \multicolumn{1}{c}{TiO$_2$} 
                                  & \multicolumn{1}{c}{SrO} &  \multicolumn{1}{c}{TiO$_2$} \\
\hline                                  
FI &  $-$10.368  &    13.468    &     6.837   &        3.100  &   $-$3.531 \\
LM &   $-$0.444  &    $-$4.934   &       5.343   &     $-$5.378  &      4.899 \\
\hline
RI &  ${-10.812}$  &     8.534     &    12.180    &     ${\bf -2.278}$  &    ${\bf 1.368}$
\end{tabular}
\end{ruledtabular}
\end{center}
\caption{Flexocoupling coefficients of a relaxed SrTiO$_3$ slab. The frozen-ion (FI),
lattice-mediated (LM) and total relaxed-ion (RI=FI+LM) values of the bulk, surface and
total slab response are reported. $\varphi^{\rm bulk}$(FI) was inferred by using
the $\varphi^{\rm bulk}_{xx,\beta \beta}$ values of Table~\ref{tab1} and the 
calculated $\nu$ parameter.
The $\varphi^{\rm surf}_{xx,{\rm eff}}$(FI) values, however, do not correspond to the linear 
combination of $\varphi^{\rm surf}_{xx,xx}$ and $\varphi^{\rm surf}_{xx,yy}$
from Table~\ref{tab1}: the $\varphi^{\rm surf}_{xx,{\rm eff}}$(FI) reported here
refer to the frozen-ion deformation of a \emph{relaxed}, rather than truncated-bulk, 
surface geometry. Volt units are used throughout. \label{tab2}}
\end{table}

To investigate the relaxed-ion response of the film, we shall consider
an ``effective'' bending deformation of the type~\cite{artgr}
\begin{eqnarray}
\varepsilon_{{\rm eff},x} &=& \varepsilon_{yy,x}- \nu \varepsilon_{xx,x}.
\label{eqeff}
\end{eqnarray}
%
The coefficient $\nu= \mathcal{C}_{xx,yy} / \mathcal{C}_{xx,yy}$ 
is the ratio of the transversal and longitudinal components of the bulk elastic tensor, $\mathcal{C}_{\alpha \beta \gamma \lambda}$,
and accounts for the mechanical equilibrium condition, i.e. it ensures that the 
stress field vanishes everywhere in the interior of the deformed sample.
(Our calculated value for SrTiO$_3$ is $\nu=0.2914$.)
As before, the overall flexocoupling coefficient of the slab can be written in
terms of a bulk and a surface contribution, 
\begin{equation}
\varphi_{xx,{\rm eff}} = \varphi_{xx,{\rm eff}}^{\rm bulk} + \varphi_{xx,{\rm eff}}^{\rm surf},
\end{equation}
whose respective impact on the induced electrostatic potential follow the 
form of Fig.~\ref{efields}(e).
The relaxed-ion (RI) value of $\varphi_{xx,{\rm eff}}^{\rm bulk}$ consists in a 
frozen-ion (FI) contribution, given in terms of the already known 
$\varphi_{xx,\beta \beta}^{\rm bulk}$ values, plus a lattice-mediated (LM) part,
which we calculate as detailed in Supplementary Note 2. The corresponding contributions
to $\varphi_{xx,{\rm eff}}^{\rm surf}$ are calculated by applying a uniform strain
of the type $\varepsilon_{{\rm eff}} = \varepsilon_{yy}- \nu \varepsilon_{xx}$
to fully relaxed slab supercells (see Supplementary Note 3).

A summary of the results is reported in Table~\ref{tab2}.
The values marked with boldface font, i.e., the flexocoupling coefficient of a 
fully relaxed SrTiO$_3$ slab subjected to bending, are the main result of this work.
[The beam-bending case 
is easily recovered by multiplying the reported values by 
$\tau =\mathcal{C}_{xx,xx} /  (\mathcal{C}_{xx,xx} + \mathcal{C}_{xx,yy})$.
By using the calculated elastic constants of bulk SrTiO$_3$, reported in 
Table S2, we find $\tau=0.77$.]
Note their substantial departure with respect to the 
corresponding bulk coefficient, confirming the dramatic impact of the 
surface structural and electronic properties on the electromechanical
response of the system. 
Remarkably, the aforementioned response coefficients are opposite in sign 
depending on whether a SrO- and TiO$_2$-terminated slab is considered.
In fact, the surface shows an even larger termination dependence
at the frozen-ion level, but with \emph{opposite} sign:
the LM contribution to $\varphi^{\rm surf}$ 
depends so strongly on the termination that its inclusion results 
in a voltage reversal, both in the TiO$_2$- and SrO-type slabs. 
(A microscopic analysis of the surface relaxations is
provided in Supplementary Note 3.)
By contrast, the LM contribution to the bulk flexocoupling coefficient
is relatively minor, about one order of magnitude smaller than any other value
reported in the table, and of little impact on the final results. 
This constitutes a substantial departure from the commonly accepted idea
that bulk lattice-mediated mechanisms are predominantly responsible for 
the flexoelectric polarization. On the contrary,
our results indicate that, by modifying the surface, one can fully control the magnitude, 
and even the sign, of the flexoelectric effect, in stark contrast with
previous assumptions.

These results have profound implications, both for the interpretation of the
experiments and for the optimization of electromechanical devices based on the
flexoelectric effect. 
In particular, achieving a control over the properties of the sample surface
appears crucial to maximizing the flexoelectric performance of a material like SrTiO$_3$.
Given the rich variety of surface structures and compositions that are accessible
to perovskite oxides depending on thermodynamic conditions and treatment procedures, 
this opens up an essentially unlimited range of opportunities for device design.

{\bf Acknowledgments.} We thankfully acknowledge the computer resources, 
technical expertise and assistance provided by the Supercomputing Center of Galicia (CESGA).

\bibliography{flexo,max-feb28}

\end{document}